**Title**

Comprehensive Quality Investigations of Wire-feed Laser Additive Manufacturing by Learning of Experimental Data

**Authors**

Sen Liu[a,b], Craig Brice[a,b*], Xiaoli Zhang[a,b*]

[*]Corresponding authors: craigabrice@mines.edu; xlzhang@mines.edu

**Affiliations**

[a] Mechanical Engineering, Colorado School of Mines, Golden, CO 80401 USA
[b] The Alliance for the Development of Additive Processing Technologies, Colorado School of Mines, Golden, CO 80401 USA

**Abstract:** Wire-feed laser additive manufacturing is an emerging fabrication technique capable of highly automated large-scale volume production that can reduce both material waste and overall cost while improving product lead times. Quality assurance is necessary for implementation into critical structural applications. However, the large number of process variables along with the cost associated with traditional trial and error methods makes this difficult. This study investigates a comprehensive quality framework based on learning from experimental data that will enable improved quality control along with consistent microstructural features of the part. Specifically, a comprehensive experimental data across multiple process variables and output characteristics in terms of overall bead quality, geometric shape (i.e. bead height, width, fusion zone depth, etc.), and microstructural features are collected. The predicted process-geometry-microstructure relations are then captured by virtue of data-driven machine learning models. The properties of printed beads are visualized based on an extensive range of processing space within a 3-dimensional contour map. The insights and impacts of process variables on bead morphology, geometric and microstructural features are comprehensively investigated for quality improvement during manufacturing processes.

# 1. Introduction

Wire-feed AM (WAM) has gained a lot of attention in recent years [1][2][3]. WAM has several advantages over other forms of metallic AM, such as high-level automation from the commercially-available robotic arm system, high volume production, and low capital costs compared to other AM methods [4][5]. WAM fabricates parts with a complex thermal history that drives directional solidification followed by multiple heating and cooling cycles. The resulting parts may have complex material microstructures and highly variable part properties [6][7][8]. Depending on the energy source used for metal deposition, WAM technologies can be classified into three groups, namely wire-feed laser AM (WLAM) [9], wire-feed arc welding AM (WAAM) [10][11][12] and wire-feed electron beam AM (WEAM) [13][14]. Because of the short laser-material interaction times and highly localized heat input, the thermal gradients and rapid solidification rates lead to a build-up of thermal stresses and non-equilibrium phases [15][1]. Non-optimal process parameters can cause molten pool instabilities, which leads to defects and geometric distortion [16][17]. Further, it has been shown that AM-specific microstructures from a rapid solidification rate significantly affect mechanical properties of the part such as strength, ductility, toughness, fatigue, and corrosion behavior [18]. Ti-6Al-4V is one of the more widely studied alloys for AM processing due to its low density, high strength, high corrosion resistance and biocompatibility [19][20]. Relatively little work has been performed to date for WLAM Ti-6Al-4V to thoroughly investigate different processing combinations that influence the geometry and microstructure features, due to the expensive data collection and the complex processing parameters interaction. To ensure optimal building conditions during WLAM for general application and adoption in industrial fields, it is critical to explore predictive process-property (PP) relations with insights of underlying physics to provide a design framework for materials with desired multi-performance functionality.

Many researchers have reported how different process parameters, such as laser power, laser travel speed, wire feed speed, and wire heat power influence Ti-6Al-4V microstructure, geometric shape, molten pool behavior, part density and surface quality [16][17][18][21][22][23]. Also, data-driven machine learning (ML) models have recently been developed to make predictions and optimize properties during AM process [24][25][26][27][28][29]. For example, an in situ quality monitoring for laser powder bed fusion (LPBF) AM was studied based on acoustic emission signals analyzed by neural networks [30][31]. A vision system was built with a high-speed camera for process image acquisition. The system detects the information from three objects: the molten pool, the vapor plume, and the spatter [32]. A process map was constructed for powder bed fusion AM using a support vector machine to classify built parts with good or bad quality [27]. A model was developed to construct maps of molten pool depth or porosity level versus process parameters for parts fabricated by LPBF AM [33][34]. The effect of weld parameters such as travel speed, accelerating voltage, and beam current on deposited layer width was studied using Taguchi method and ANOVA feature analysis [14]. The influence and correlation between processing parameters on characteristic geometry features such as bead height, width and dilution for micro laser metal wire deposition are discussed [5][9].

Although researchers have already studied the effects of process variables on bead characteristics to establish an optimal process window, most of these presented results are not comprehensive due to the high dimensionality of PP relations, each process variable and property are often investigated individually. Few publications have reported a comprehensive quality design framework for the WLAM process considering various processing variables against multiple characteristics such as bead geometric shape and microstructures, especially with data-driven ML approaches. One reason is massive resource consumption in that a large number of experiments or computationally intense simulations are often required for modeling complex PP correlations. Besides, there is a lack of domain expertise for comprehensively labeling experimental data. The appropriate statistical analysis application on experimental data requires the cooperation of expert knowledge in both data science and materials science. In all, the great challenge for PP relations modeling of WLAM is to collect a comprehensive database that consists of multiple process variables and multiple properties/characteristics under a set of controlled experiments. The underlying knowledge from the database can then be used for comprehensive quality quantification and control during WLAM process.

In this work, we collected a database from single-layer deposition experiments under a set of controlled processing parametric combinations. Bead geometry (bead height, bead width, fusion zone depth, fusion zone area, dilution and aspect ratio), overall bead quality ("failed", "rippled" or "smooth" bead morphology), and microstructures (Alpha ($\alpha$) phase thickness and Beta ($\beta$) grain length) are measured for correlation with the processing conditions. In total, this study examines about 200 samples, which significantly exceed previous wire-feed AM related publications to the authors' knowledge. The effects of processing parameters on output property and their correlations are performed with Mutual Information score and Pearson correlation matrix. The process parameters are down-selected for data-driven ML modeling to capture the complex bead process – geometry – microstructure relationships. The overall bead surface quality is formulated as a Bayesian classification model to distinguish the failed, rippled, and smooth bead morphology. The bead geometric and microstructural properties are predicted with Gaussian process regression models. A material design prediction methodology is proposed and visualize the predictions in 3D contour maps. The interpretation and insight of processing variables on the characteristics of the bead are discussed as means to improve deposited part quality.

## 2. Materials and Methods

### 2.1 Overview of workflow

The material design strategy for comprehensive quality assurance is shown in Fig. 1. It presents four main parts: a) data collection and preprocessing, b) process settings importance and correlations, c) data-driven ML modeling, and (d) predictions and insight.

The Ti-6Al-4V alloy was chosen for the WLAM process. Single layer deposits were made using a set of controlled process parameter combinations. The overall bead quality is labeled as failed, rippled, or smooth based on bead morphology. The bead geometry in cross-section is measured as bead height, bead width, fusion zone depth, and fusion zone area. The bead microstructural properties that were measured were $\alpha$ phase lath thickness and prior $\beta$ grain length in parallel and perpendicular to scan direction. Data processing includes extracting and tabulating unstructured data into a structured database. The significant outliers are removed to clean the datasets. Then both process parameters and bead properties are checked with consistent units. The experimental data are visualized with statistical process-property and property – property scatterplots to assess the range and distribution of process and property, as well as their mutual correlations as of interest. The prepared high-quality experimental data are then used as the input for subsequent process setting analysis and modeling.

The process settings importance score and setting pairs correlation are calculated and analyzed. The input data is normalized to 0~1 to reduce the implicit bias introduced from differences in data scales and ranges. In the collected database, 46 processing parameters are recorded; whereas some of these settings are not associated with or contribute little to the output property. The process settings down-selection process can reduce model complexity and make the model more accurate and interpretable.

The process – property relations are then modeled with data-driven classification and regression models. Prior to any internal property, the first consideration should be that the printed beads have a smooth morphology, and lumpy or intermittent beads should be avoided. The classification model is therefore applied to bead surface quality and overall appearance. The quantitative bead characteristics such as bead geometric shape and microstructural features are modeled with predictive regression models. Once the models are trained using the known process parameters, they can then predict characteristics of the printed part.

The output characteristic in terms of overall bead quality, geometric shape, and microstructure can be predicted with the verified ML models. Specifically, a set of 3-dimensional (3D) contour maps are created to provide insight and interpretation of PP relations.

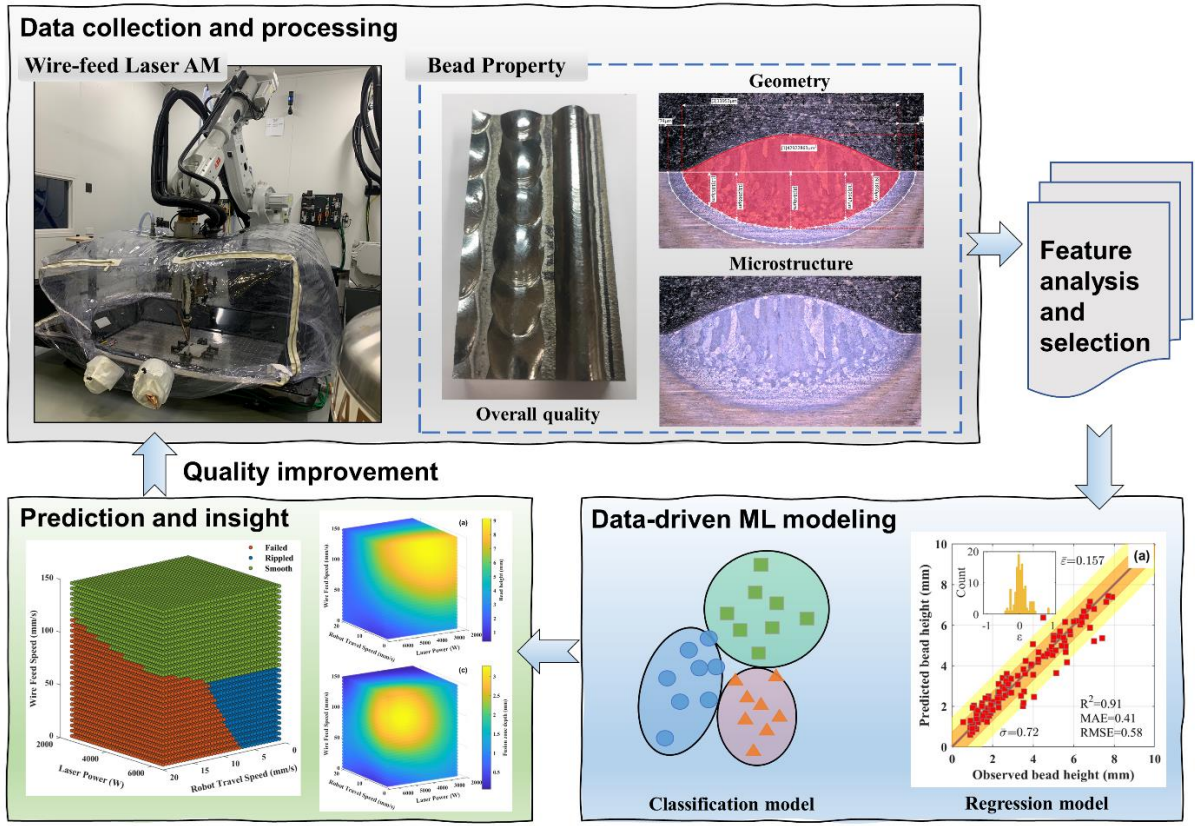

**Figure 1.** The comprehensive quality assurance framework for material design is demonstrated on WLAM Ti-6Al-4V. The experimental data were collected from single-layer deposits under a set of controlled process parametric combinations. Bead overall quality, geometric shape, and microstructure are experimental measured. Data-driven modeling includes ML classification and regression models, which established the correlations between process variables and output characteristics. The comprehensive investigations of the printed bead are presented as the interpretation and insight of process – geometry – microstructure correlations.

## 2.2 Experimental data collection across multiple properties

The experiments are conducted under a set of controlled process parameters on a WLAM system, as shown in Fig. 2(a). A 6kW laser head and a wire-feed system are mounted to the robot arm. The robot arm and deposition area are covered in a closed transparent chamber filled with argon gas to protect the material during operation. The feedstock material is 1.59mm diameter Ti-6Al-4V wire. A set of sensors including an optical camera, multiple pyrometers, a spectrometer, and an acoustic sensor are installed in the chamber to monitor the visual, thermal, and positional information of molten pool during manufacturing in real-time. The WLAM process parameters are recorded during manufacturing as listed in Table 1. In total, it includes eight main process categories with total forty-six process parameters. The single beads of Ti-6Al-4V are deposited and analyzed without post heat-treatments. In total, 179 experimental beads are deposited under different process parameter combinations for data-driven modeling and analysis. These datasets are tabulated and illustrated in Table S1.

The deposited bead surface is qualitatively characterized. No heat treatment is performed on any of the bead samples produced in this study. The overall bead surface quality can be labeled by domain expertise with failed, rippled and smooth, as illustrated in Fig. 2(b). The "failed" bead indicates the intermittent, rough, or uneven bead morphology, while "smooth" bead is continuous, smooth, and well-distributed without significant bead waves appeared. The "rippled" bead is a medium state that continuous, well-distributed, but rough with wave appearance.

The printed bead is then sectioned, mounted, and polished to identify the bead geometry in cross-section. These specimens were first prepared using standard metallographic techniques. bead height, bead width, fusion zone depth, and area is performed using a Keyence VHX-5000 optical microscope. The dimensional and microstructural information of bead in cross-section are shown in Fig. 2(c). The samples are then etched for 40–60 s in Kroll reagent (1 ml HF+2 ml $HNO_3$+50 ml $H_2O$) to reveal the grain structure. Optical microscopy under polarized light is then used to distinguish the prior-$\beta$ grains by their crystal orientation and $\alpha$ lath thickness. The microstructural image of $\alpha$ lath and prior-$\beta$ grains are analyzed using ImageJ software to obtain statistical information.

**Table 1.** The category and descriptions of process settings that are recorded during the WLAM manufacturing process.

| Process category | Process index | Process description | Process category | Process index | Process description |
|---|---|---|---|---|---|
| **Laser Equipment Data** | 1 | Optical Channel | **Print Fill Stage Data** | 25 | Fill Time (s) |
| **Post-Print Data** | 2 | Burn Back Time (s) | | 26 | Heat Wire-feed Speed (mm/s) |
| | 3 | Gas Off Delay (s) | | 27 | Laser Power (W) |
| | 4 | Laser Off Delay (s) | | 28 | Wire Power (kW) |
| | 5 | Pull Out Dist (mm) | | 29 | Wobbler Amplitude (mm) |
| | 6 | Roll Back Time (s) | | 30 | Wobbler Frequency (Hz) |
| **Pre-Print Data** | 7 | Gas Pre-Flow Time (s) | **Print Ignition Stage Data** | 31 | Laser Power (W) |
| | 8 | Gas Purge Time (s) | | | |
| | 9 | Pre-Time Recorded (s) | | | |
| **Process Summary** | 10 | Ending O2 (ppm) | | 32 | Laser Pre-Time (s) |
| | 11 | Laser Power (W) | | 33 | Laser Start Move Delay (s) |
| | 12 | Measured Pre-Time (s) | | 34 | Weld Power Source/Hotwire Mode |
| | 13 | Measured Weld Time (s) | | 35 | Wire Power (kW) |
| | 14 | Robot Travel Speed (mm/s) | | 36 | Wobbler Amplitude (mm) |
| | 15 | Starting O2 (ppm) | | 37 | Wobbler Frequency (Hz) |
| | 16 | Total Length (mm) | **Print Main Stage Data** | 38 | Heat Wirefeed Speed (mm/s) |
| | 17 | Wire Feed Speed (mm/s) | | 39 | Laser Power (W) |
| | 18 | Wire Heat Power (kW) | | 40 | Travel Speed (mm/s) |
| | 19 | Heat Wire feed Speed (mm/s) | | 41 | Weave Length (mm) |
| | 20 | Laser Power (W) | | 42 | Weave Width (mm) |

| | 21 | Weld Power Source/Hot-wire Mode | | 43 | Weld Power Source/Hotwire Mode |
| --- | --- | --- | --- | --- | --- |
| **Print End Stage Data** | 22 | Wire Power (kW) | | 44 | Wire Power (kW) |
| | 23 | Wobbler Amplitude (mm) | | 45 | Wobbler Amplitude (mm) |
| | 24 | Wobbler Frequency (Hz) | | 46 | Wobbler Frequency (Hz) |

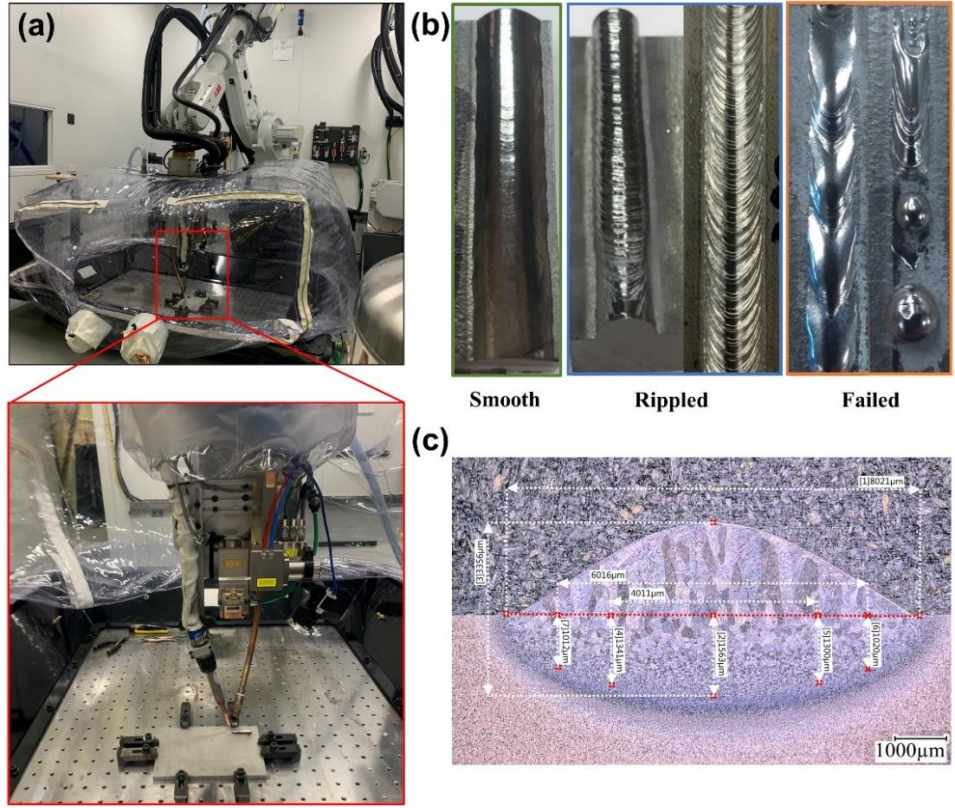

**Figure 2.** The experiments data collection from (a) WLAM system under a set of controlled process parameters, (b) bead overall quality is labeled as smooth, rippled, and failed categories, and (c) bead geometry features and characteristics of microstructures.

**2.3 Process settings analysis and down-selection**

Among all process-related settings that have been generated during WLAM, some are not directly associated with output properties. The down-selection process can reduce model complexity and make the model more interpretable. There are 46 raw input process settings monitored and recorded during manufacturing (Table 1). Before final tuning and verification of the ML models, it is desirable to analyze the inputs for redundancy and/or insignificance, such that the number of inputs may be reduced, in turn reducing the functional complexity of the ML models. Here, the relative importance of the process settings in determining bead properties is ranked using the mutual information (MI) score method [35][36]. It equals zero if and only if the setting feature is independent of the output property, and a higher MI score larger than 0.3 means higher importance on property. Process settings' feature correlation and redundancy are evaluated using Pearson correlation [37]. Process setting pairs with a correlation coefficient larger than 0.90 are highly correlated. The Scikit-learn python implementation of these algorithms was used [38].

**2.4 Machine learning models for process-property correlation**

**2.4.1 Overall bead quality model**

For classification of bead overall quality (e.g. smooth, rippled, and failed) from process setting parameters, many ML methods can be used for inference such as Naïve Bayes (NB) networks, logistic regression (LR), support vector machine (SVM), k-nearest neighbor (KNN), artificial/deep neural networks (ANN/DNN), and random forests (RF) [39]. Among them, NB networks have proven to be effective in the process – property modeling for laser powder bed fusion AM in our previous paper [40]. A NB classifier is a conditional probability model. Given a problem instance represented by $n$ input process settings, $X = (x_1, ..., x_n)$, the classifier assigns to the $k^{th}$ property level $C_k$ with probability $p(C_k|x_1, ..., x_n)$. Using Bayes' theorem, the probability that a certain property level $C_k$ occurs under process setting status $X$, the conditional probability, can be related to the conditional likelihood of $X$, the prior of $C_k$, and the marginal distribution of $X$,

$$p(C_k|X) = \frac{p(X|C_k)p(C_k)}{p(X)} \qquad (1)$$

To effectively build an inference NB model that can determine the part overall quality level $C_k$ at given processing conditions $X$, the maximum a posterior (MAP) decision rule is used to estimate property $\widehat{C_k}$, the most likely value for property level $k$, which is defined as,

$$\widehat{C_k} = \underset{k}{\mathrm{argmax}}\, p(C_k) \prod_{i=1}^{n} p(x_i|C_k). \qquad (2)$$

Where $p(C_k)$ is the prior probability of property level $k$. $p(X|C_k)$ is conditional likelihood probability, which indicates the probability of process status $X$ occurs at a given property observation.

**2.4.2 Bead geometry and microstructure models**

A predictive regression model is built to estimate the continuous value of bead geometric and microstructural characteristics from input process parameters. The ML algorithms such as support vector regression (SVR), random forest (RF), Gaussian process regression (GPR) can be used for continuous property value prediction [41]. GPR model has shown to be effective in modeling compositions – process – property correlations [42]. Here, we choose GPR algorithms for bead geometric and microstructural features prediction because of GPR model can not only estimate the response with superior prediction accuracy, but also the standard deviation of the response made this model more suitable for extrapolation near the supported domain. Knowing the process setting parameters $X$, property observations $Y$ and optimal hyper-parameters $\theta$, the property prediction $Y_*$ given a specific unknown process settings $X_*$ is given by,

$$Y_*|\mathbf{X},\mathbf{Y},X_* \sim N(\hat{Y}_*, \sigma_{\hat{Y}}^2(X_*)) \tag{3}$$

where the mean of prediction $\hat{Y}_* = K(X_*,\mathbf{X})[K(\mathbf{X},\mathbf{X}) + \sigma_n^2 I]^{-1}\mathbf{Y}$, and the prediction variance/uncertainty $\sigma_{\hat{Y}}^2(X_*) = K(X_*,X_*) - K(X_*,\mathbf{X})[K(\mathbf{X},\mathbf{X}) + \sigma_n^2 I]^{-1}K(\mathbf{X},X_*)$. Thus, the prediction at unknown process settings $X_*$ is given as a normalized distribution with mean $\hat{Y}_*$ and its associated variance $\sigma_{\hat{Y}}^2$. The covariance function $K(x_i, x_j)$ captures the dependence between different locations $x_i$ and $x_j$ within the feature space. The isotropic squared exponential covariance function was used. $\boldsymbol{\theta} = \{\sigma_n^2, \sigma_f^2, l\}$ denote the GPR model hyper-parameters needed to be calculated based on the observations datasets $\{\mathbf{X}, \mathbf{Y}\}$. Overfitting and underfitting of the GPR models are avoided by using the conjugate gradients method to tune hyper-parameters, considering bias-variance tradeoff. More specifically, the high bias region means the under-fitting of model and high variance region indicates over-fitting. The boundary of under-fitting and over-fitting region presents the optimal hyper-parameters. GPR models are optimized by restarting the model training at different locations in the hyperparameter space. Matlab GPML version 4.2 [43] was used to implement the GPR models.

For the classification model of printed bead quality levels, confusion matrix and receiver operating characteristic (ROC) curve are used to evaluate the performance of the model [40][44][45][46]. The error of the regression model for bead geometric and microstructural characteristics uses metrics such as coefficient of determination ($R^2$), root-mean-square error (RMSE), mean absolute error (MAE), and mean relative error ($\bar{\varepsilon}$). Both of classification and regression models are quantified with ten-fold cross-validation to evaluate the model prediction performance out-of-sample training [40][47].

## 2.5 Property prediction methodology

To demonstrate the potential of the model for property design, the model was applied to predict a property of a printed bead under various process combinations, which have not been experimentally studied. According to process settings analysis in Section 3.2 and domain knowledge for parameter control of WLAM, the process settings laser power (LP), robot travel speed (RTS), and wire feed speed (WFS) are the most dominant and easily adjustable parameters for bead quality control. Thus, to better visualize the prediction in lower 3D feature space, LP, RTS, and WFS are chosen as changing process variables, whereas other settings such as wire heat power and oxygen content are kept as constant, in which we set them as the average value of experiments in the database. The prediction confined to three process settings is also for better visualization and insight from 3D contour maps. However, the developed framework can generalize to any process combinations of the model input process settings for property prediction and process window optimization. The prediction maps of property against the variation of process parameters are usually drawn with 2-dimensional curves or maps, due to the limited experimental data or process parameter constrict on a specific range [3][9][14][17][27][48]. Compared with prior related studies, our prediction methodology generates a large range of process space in a 3D contour space, which benefits the visualization for complex process interaction or coupling effect for the property outcome, comprehensive quality investigation, and significantly reduces the prediction figures needed.

The constituent process space is confined as $2500 \leq LP \leq 6500$ W, $1 \leq RTS \leq 20$ mm/s, and $5 \leq WFS \leq 150$ mm/s, as the region of interest. Their variation step size are 100 W, 0.5 mm/s, and 5 mm/s, respectively. Other process settings are set as constant while prediction, such as ending oxygen content 23 ppm, measured weld-time 33 s, and wire heat power 0.6 kW. Therefore, it exhaustively explored the 47,970-point design space to estimate the bead printed property using the trained ML models. 3D contour maps visualizing the resulting calculations are given in Fig. (10-13) for predictions $\mu$ of bead quality, geometry, and microstructures, and Fig. S(1-3) for the predicted probability or uncertainty, $\sigma$. Finally, as given in Fig. (14,15), we studied the effects of wire heat power for the predictions of bead overall quality and geometric shape at a higher wire heat power 0.9 kW. The 3D predictions in the contour maps can also be mined for physical insights and instruct manufacturing parameterizations.

## 3. Results and Discussions

Specifically, in Section 3.1, the experimental data are visualized and assessed by process- property scatter plot for ML modeling. The process settings' importance score for the property, the process setting pairs' correlations, and down-selection process are performed in Section 3.2. The process – property relations are modeled by machine learning from experimental data and validated with error metrics in Section 3.3. A property prediction methodology is used to visualize the property prediction and insight for the complex process – property relations in Section 3.4.

## 3.1 Assessment of the database

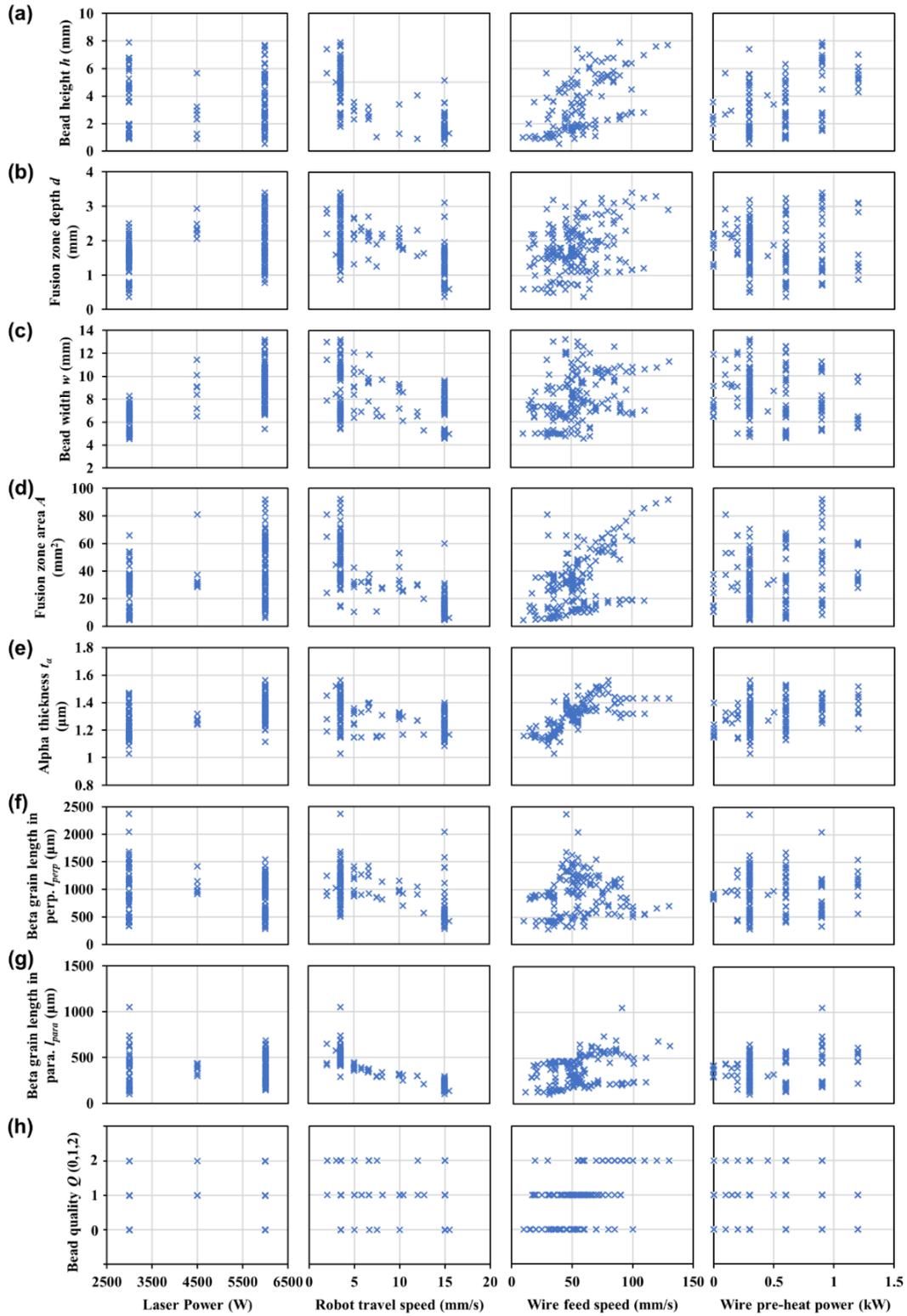

**Figure 3.** Visualization of the database for process – property relations. The scatter plot of process variables versus properties across (a) bead height, (b) fusion zone depth, (c) bead width, (d) fusion

zone area, (e) α lath thickness, (f) β grain length in perpendicular, (g) β grain length in parallel to scan direction, and (h) bead overall quality (0: failed, 1: rippled, 2: smooth).

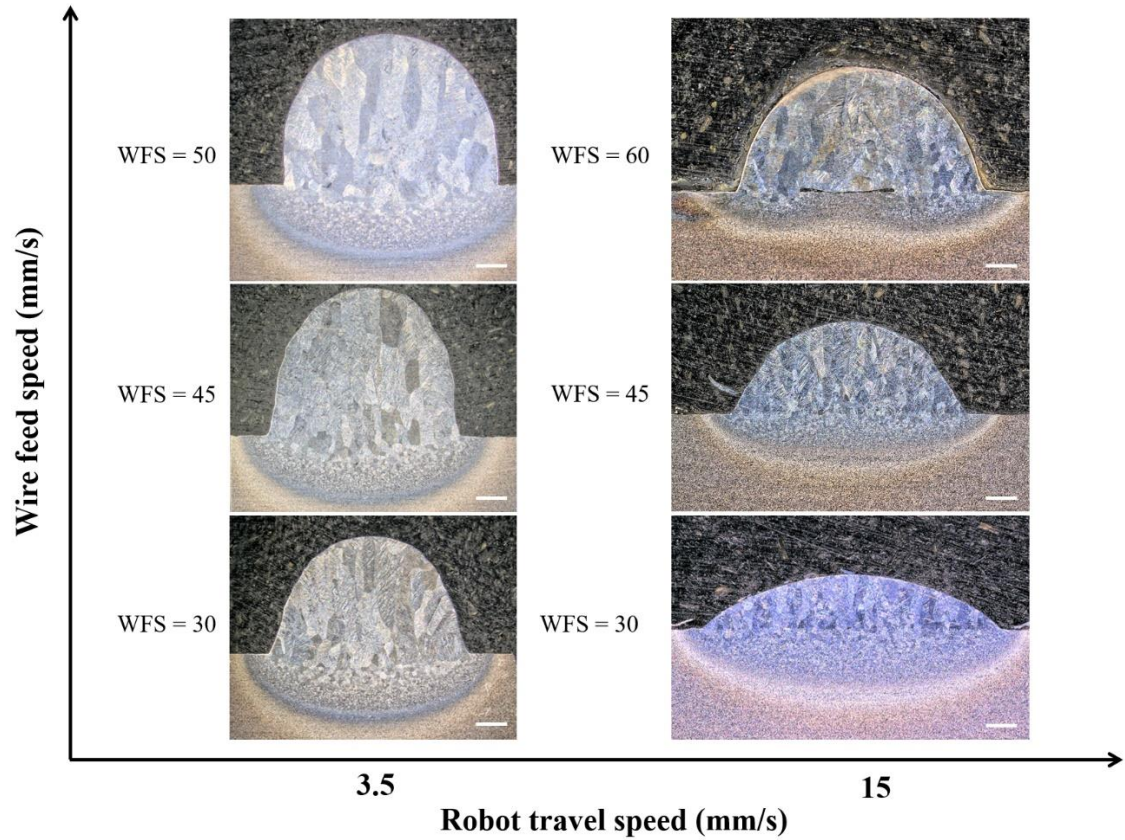

**Figure 4.** Visualization of bead geometry variations against the changes of wire feed speed and robot travel speed. The laser power is set as 3000 W and wire pre-heat power is 0.6 kW. The scale bar in the image indicates 1000 μm.

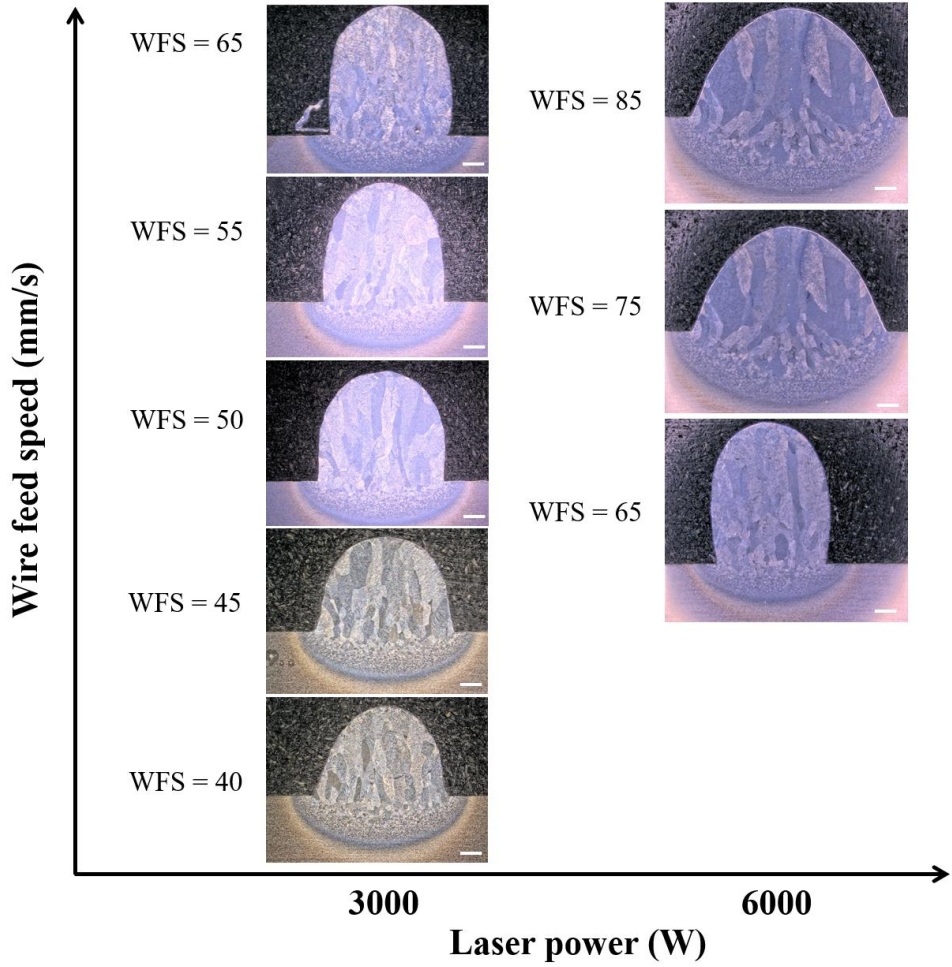

**Figure 5.** Visualization of bead geometry variations against the changes of laser power and wire feed speed. The robot travel speed is set as 3.5 mm/s and wire pre-heat power is 1.2 kW. The scale bar in the image indicates 1000 μm.

The datasets for representing process – property relations have been displayed in a scatter plot, as shown in Fig. 3. Since 46 process settings are involved in WLAM, only laser power, robot travel speed, wire feed speed, and wire heat power are selected for visualization since they have been determined to be the most significant based on acquired domain knowledge by process experts. The properties visualized include bead height ($h$), fusion zone depth ($d$), bead width ($w$), fusion zone area ($A$), $\alpha$ phase thickness ($t_\alpha$), $\beta$ grain length in parallel to scan direction ($l_{para}$), $\beta$ grain length in perpendicular to scan direction ($l_{perp}$), and bead overall quality ($Q$). Overall bead quality was indicated as 1: failed, 2: rippled, and 3: smooth. In general, it reveals both typical and atypical values for adjustable parameters for printing Ti-6Al-4V: such as laser power range (2500~6500 W), robot travel speed (approximately 16 mm/s), wire feed speed (up to 130 mm/s) and wire pre-heat power (0~1.2 kW). It observed that bead geometry features such as $h$, $w$, and $A$ decrease with robot travel speed. $h$ and $A$ tend to increase with WFS. Concerning microstructures, $\alpha$ lath thickness and $β$ grain length tend to decrease with the increase of RTS. The $α$ lath thickness is likely to increase with LP and WFS. But the precise property variations value and rate are difficult to capture through the scatter plot due to the complex process coupling effects and multiple geometry characteristics coupling effects. As a case demonstration, visualization of bead incision geometry variations against the changes of WFS, RTS and LP are presented in Fig. 4 and Fig. 5. They show a complex and nonlinear bead geometry variation trend even with only two processing variables. Many other process – property relations exist but cannot singularly be revealed by scatter plot and microscopy image visualization, indicating that the relationship correlations are of a higher order. A data-driven ML approach will be suitable for modeling a large amount of experimental data and visualizing these complex process – property relationships within the datasets.

### 3.2 Process settings importance and correlation

The significance of input parameters on deposited bead property based on MI score is shown in Table 2. The process settings' importance score threshold is set as 0.3, and those settings with a score less than 0.3 are removed. To make input settings consistent for regression models of bead geometry and microstructural properties, the average value of $h$, $A$, $w$, $d$, $t_a$, $l_{para}$ and $l_{perp}$ are used for the selection of significant process settings for these regression models. The quality classification model $Q$ is shown in the last column of the table for settings selection. The settings with high MI scores are highlighted in red in the last two columns of the Table 2. It can be seen many process settings in post-print, pre-print, print end stage, print fill stage, and print ignition stage setting categories exhibit low MI scores and are removed for subsequent modeling. The process setting categories of process summary and print main stage are generally with high setting importance, such as process setting 10: laser power, 14: robot travel speed, 17: wire feed speed, 18: wire power, etc.

After the process settings importance ranking, there remain many process settings as highlighted in red of Table 2. Eliminating the strongly correlated setting pairs will reduce the total number of input setting features. The heatmap matrix of process settings for this purpose is presented in Fig. 6(a). The setting pairs with a correlation coefficient larger than 0.90 are taken to be highly correlated. The most important (assessed via MI scores) variables among the highly correlated setting pairs are retained. Hence, redundant settings 38, 39, 40, and 44 are removed since their higher correlation with settings 17, 11, 14, and 18. The retained process settings for the regression model are shown in Fig. 6(b). In the same way, redundant settings 9, 30, 38, and 40 are removed for the classification model, and retained settings are shown in Fig. 6(c). In the retained process settings for ML modeling, robot travel speed, laser power, wire feed speed, wire heat power and ending oxygen content are significant process settings for bead output properties as expected. Among them, robot travel speed is the most significant process settings, whereas laser power is low relative to robot travel speed, likely due to insufficient statistical laser power sampling, as shown in Fig. 3.

**Table 2.** The process setting down-selection based on mutual information (MI) score, which indicates the overall significance of process settings in determining each output property. The specific process name is labeled as an index as indicated in Table 1. To make input settings consistent for regression models of bead geometry and microstructural properties, average value of $h$, $A$, $w$, $d$, $t_a$, $l_{para}$ and $l_{perp}$ is used for selection of significant process settings for the regression model. The quality classification model $Q$ is shown in the last column of the table. The settings importance score threshold is 0.3, the highly important process settings for regression models and classification model $Q$ are shown in red.

| Process category | Process index | MI score of property | | | | | | | | |
|---|---|---|---|---|---|---|---|---|---|---|
| | | $h$ | $A$ | $w$ | $d$ | $t_a$ | $l_{para}$ | $l_{perp}$ | *Average* | $Q$ |
| Laser Data | 1 | 0.00 | 0.00 | 0.00 | 0.00 | 0.00 | 0.02 | 0.00 | 0.00 | 0.00 |
| | 2 | 0.00 | 0.00 | 0.00 | 0.00 | 0.00 | 0.01 | 0.00 | 0.00 | 0.00 |
| Post-Print Data | 3 | 0.00 | 0.00 | 0.00 | 0.00 | 0.00 | 0.00 | 0.00 | 0.00 | 0.25 |
| | 4 | 0.26 | 0.40 | 0.17 | 0.23 | 0.11 | 0.07 | 0.18 | 0.20 | 0.00 |
| | 5 | 0.16 | 0.23 | 0.03 | 0.22 | 0.06 | 0.04 | 0.11 | 0.12 | 0.07 |
| | 6 | 0.12 | 0.16 | 0.14 | 0.61 | 0.13 | 0.10 | 0.24 | 0.22 | 0.09 |
| Pre-Print Data | 7 | 0.00 | 0.01 | 0.00 | 0.08 | 0.05 | 0.00 | 0.00 | 0.02 | 0.00 |
| | 8 | 0.00 | 0.00 | 0.00 | 0.00 | 0.00 | 0.00 | 0.00 | 0.00 | 0.00 |
| | 9 | 0.40 | 0.29 | 0.17 | 0.44 | 0.07 | 0.19 | 0.18 | 0.25 | *0.46* |
| Process Summary | 10 | 0.42 | 0.34 | 0.48 | 0.62 | 0.10 | 0.23 | 0.16 | *0.33* | *0.35* |
| | 11 | 0.47 | 0.39 | 0.97 | 0.61 | 0.29 | 0.22 | 0.22 | *0.45* | *0.48* |
| | 12 | 0.41 | 0.28 | 0.18 | 0.45 | 0.07 | 0.18 | 0.18 | 0.25 | *0.63* |
| | 13 | 0.93 | 0.63 | 0.67 | 0.52 | 0.22 | 0.43 | 0.54 | *0.56* | *0.77* |
| | 14 | 0.92 | 0.97 | 0.97 | 0.91 | 0.36 | 0.76 | 1.00 | *0.84* | *0.90* |
| | 15 | 0.31 | 0.31 | 0.24 | 0.39 | 0.11 | 0.07 | 0.08 | 0.22 | *0.41* |
| | 16 | 0.41 | 0.36 | 0.00 | 0.22 | 0.19 | 0.36 | 0.15 | 0.24 | *0.61* |
| | 17 | 0.50 | 0.66 | 0.89 | 0.10 | 1.00 | 1.00 | 0.65 | *0.69* | *0.54* |
| | 18 | 0.66 | 0.29 | 0.49 | 0.37 | 0.16 | 0.47 | 0.28 | *0.39* | *0.30* |
| Print End Stage Data | 19 | 0.45 | 0.33 | 0.58 | 0.82 | 0.13 | 0.34 | 0.32 | *0.42* | 0.00 |
| | 20 | 0.00 | 0.02 | 0.00 | 0.17 | 0.00 | 0.00 | 0.00 | 0.03 | 0.13 |
| | 21 | 0.00 | 0.00 | 0.03 | 0.07 | 0.00 | 0.04 | 0.01 | 0.02 | 0.00 |
| | 22 | 0.01 | 0.06 | 0.15 | 0.13 | 0.06 | 0.19 | 0.08 | 0.10 | 0.00 |
| | 23 | 0.00 | 0.00 | 0.00 | 0.00 | 0.00 | 0.00 | 0.00 | 0.00 | 0.00 |
| | 24 | 0.00 | 0.13 | 0.00 | 0.20 | 0.09 | 0.02 | 0.00 | 0.06 | 0.00 |
| Print Fill Stage Data | 25 | 0.18 | 0.17 | 0.12 | 0.35 | 0.03 | 0.04 | 0.08 | 0.14 | 0.00 |
| | 26 | 0.04 | 0.14 | 0.31 | 0.46 | 0.11 | 0.23 | 0.16 | 0.21 | 0.00 |
| | 27 | 0.00 | 0.00 | 0.00 | 0.00 | 0.04 | 0.00 | 0.00 | 0.01 | 0.00 |
| | 28 | 0.02 | 0.03 | 0.13 | 0.16 | 0.05 | 0.17 | 0.08 | 0.09 | 0.15 |
| | 29 | 0.00 | 0.00 | 0.00 | 0.04 | 0.01 | 0.00 | 0.00 | 0.01 | 0.00 |
| | 30 | 0.00 | 0.14 | 0.00 | 0.11 | 0.07 | 0.05 | 0.01 | 0.05 | *0.32* |
| Print Ignition Stage Data | 31 | 0.13 | 0.02 | 0.07 | 0.25 | 0.00 | 0.11 | 0.09 | 0.09 | 0.00 |
| | 32 | 0.23 | 0.28 | 0.05 | 0.34 | 0.06 | 0.13 | 0.20 | 0.19 | 0.03 |
| | 33 | 0.00 | 0.00 | 0.00 | 0.00 | 0.00 | 0.00 | 0.00 | 0.00 | 0.00 |
| | 34 | 0.00 | 0.01 | 0.00 | 0.00 | 0.00 | 0.00 | 0.00 | 0.00 | 0.09 |
| | 35 | 0.32 | 0.15 | 0.44 | 0.41 | 0.17 | 0.22 | 0.25 | 0.28 | 0.05 |
| | 36 | 0.00 | 0.00 | 0.00 | 0.00 | 0.01 | 0.06 | 0.00 | 0.01 | 0.08 |
| | 37 | 0.00 | 0.13 | 0.00 | 0.19 | 0.06 | 0.03 | 0.00 | 0.06 | 0.00 |
| Print Main Stage Data | 38 | 0.53 | 0.65 | 0.91 | 0.12 | 0.97 | 1.00 | 0.65 | *0.69* | *0.46* |
| | 39 | 0.47 | 0.38 | 1.00 | 0.63 | 0.30 | 0.21 | 0.22 | *0.46* | 0.28 |
| | 40 | 1.00 | 1.00 | 0.96 | 1.00 | 0.36 | 0.79 | 0.99 | *0.87* | *1.00* |
| | 41 | 0.00 | 0.01 | 0.01 | 0.11 | 0.00 | 0.03 | 0.00 | 0.02 | 0.11 |
| | 42 | 0.00 | 0.00 | 0.00 | 0.00 | 0.00 | 0.00 | 0.00 | 0.00 | 0.00 |

| | | | | | | | | | |
|---|---|---|---|---|---|---|---|---|---|
| | 43 | 0.00 | 0.01 | 0.00 | 0.14 | 0.02 | 0.00 | 0.07 | 0.03 | 0.00 |
| | 44 | 0.66 | 0.28 | 0.49 | 0.43 | 0.16 | 0.47 | 0.29 | <span style="color:red">0.40</span> | 0.00 |
| | 45 | 0.00 | 0.00 | 0.00 | 0.06 | 0.00 | 0.00 | 0.00 | 0.01 | 0.00 |
| | 46 | 0.00 | 0.12 | 0.00 | 0.16 | 0.05 | 0.01 | 0.01 | 0.05 | 0.00 |

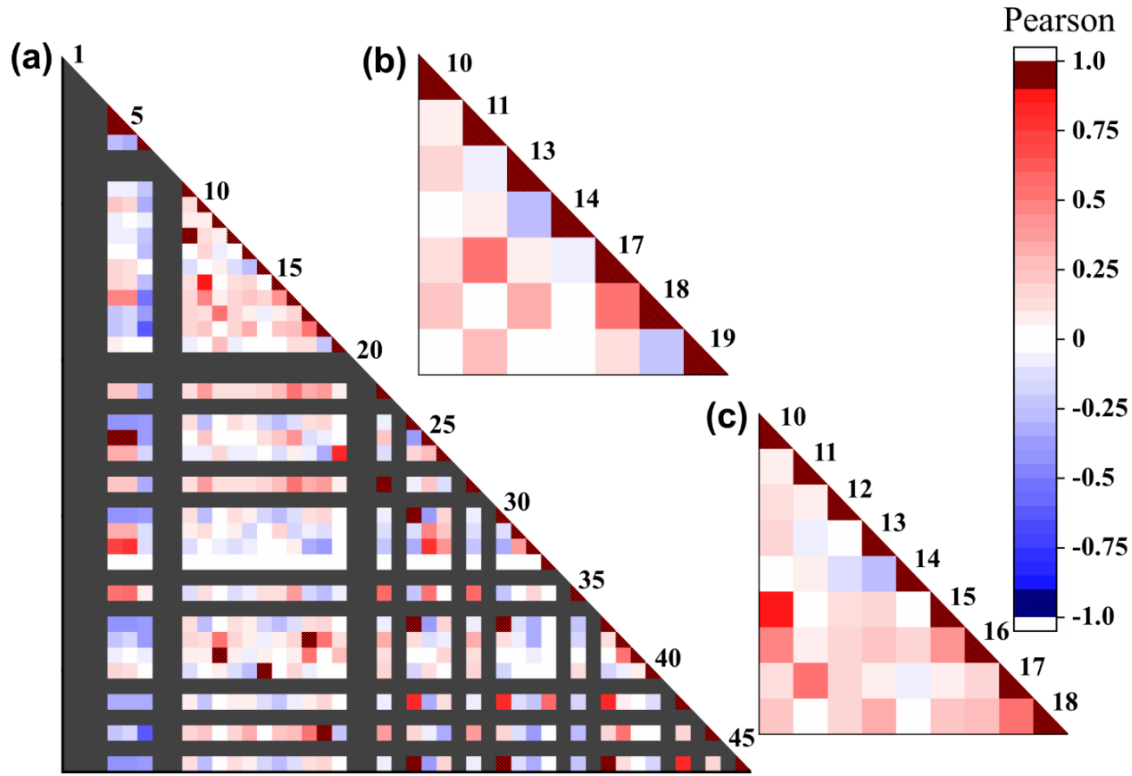

**Figure 6.** Process settings down-selection based on Pearson cross-correlation matrices indicates the relative redundancies of (a) all process settings, (b) the down-selected set of process settings used for regression models, and (c) the down-selected set of process settings for bead overall quality classification model, by considering the MI score and Pearson coefficient. The setting pairs with a correlation coefficient larger than 0.90 are taken to be highly correlated, which is covered with the black grid in heatmaps. The grey solid column or rows means NAN values since these settings are constant values in raw datasets.

### 3.3 Quantitative assessments of data-driven modelling

### 3.3.1 Bead overall quality model

The overall bead quality can be labeled as failed, rippled, and smooth as indicated in Section 2.2. In the NB classification model of bead quality, 0 indicates failed, 1 for rippled, and 2 for smooth bead quality. In the confusion matrix of Fig. 7(a), out of 179 observed datasets, the model correctly predicts 132 and misclassifies 47 with an overall prediction accuracy of 73.74%. Specifically, the 53 actual datasets with failed quality property in the first row, the model predicted 43 correctly and only predicted 10 with rippled and smooth quality; the prediction accuracy is 81.13%.

And in 94 observed datasets with rippled quality property in the second row, it predicted 71 correctly with rippled quality and misclassified 24 datasets; the prediction accuracy is 75.53%. In the third row, with 32 actual datasets with smooth bead quality, the NB classifier predicted 18 datasets correctly with 56.26% accuracy. Further, we use ten-fold cross-validation to estimate an out-of-sample model error of 28% for predicting bead quality levels. In a ROC curve, (0,1) represents a model whose false positive rate of zero and a true positive rate of one make it a perfect classifier. The diagonal line (Random guessing) in black divides the ROC space. The curve more towards the left top corner with a larger area under the curve (AUC) would be more desirable. The macro average AUC for the NB model is 0.83. The AUC value 0.75 of the rippled class is relatively lower than the other two classes 0.88 and 0.87, which is likely due to the imbalance of datasets. The model accuracy metrics are acceptable, and the classifier does capture the process settings - bead quality relations based on the collected experiments database.

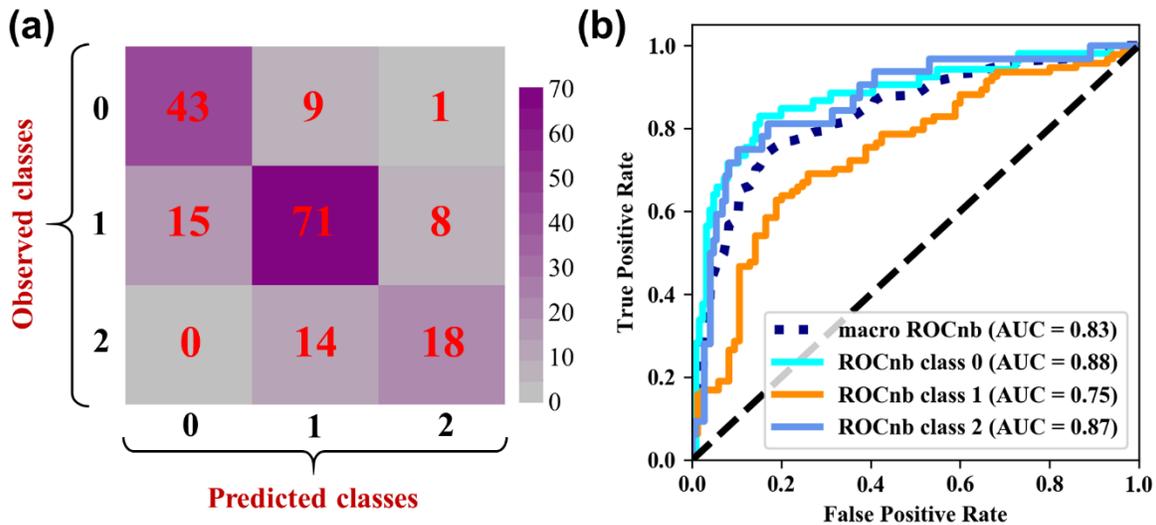

**Figure 7**. The bead overall quality classification model performance metrics with (a) confusion matrix, and (b) ROC curves of classification model, in which the ROC curve of each class and average are presented.

### 3.3.2 Bead geometry models

The geometry model performance fitted on selected process settings is shown in Fig. 8(a-d). An ideal model would place all predicted values on the brown diagonal line. Bead height is a significant indicator of bead geometry shape and was studied by many researchers in the field of wire-feed AM [3][10]. For bead height prediction, the model performs reasonably well as shown in Fig. 8(a). The $R^2=0.91$ indicates the model prediction generally aligns with the actual observations of datasets. The insert histogram shows the model predicts the datasets with mean relative error $\bar{\varepsilon} = 0.16$. The insert text shows mean predicted standard uncertainty $\bar{\sigma}$ is 0.73 mm as indicated by the orange diagonal band. The difference between the experimental observed and model-predicted height value lay along with the diagonal line within the yellow $\pm 2\bar{\sigma}$ band. The MAE and RMSE of prediction are 0.41 mm and 0.59 mm, respectively.

Bead width can be easily measured from the deposited bead. It strongly correlates with process variables and higher laser energy is likely to result in a wider deposited bead [14][49]. As in Fig. 8(b), the model shows great prediction performance for bead width with $R^2$ value 0.94. The mean error $\bar{\varepsilon}$ and uncertainty $\bar{\sigma}$ are only 0.04 and 0.68 mm, respectively. The error histogram shows the sharpest peak around 0, with over 90% of data with a predictive error lower than $\pm 10\%$.

The bead fusion zone depth is predicted in Fig. 8(c). Fusion zone depth is defined as the distance from the substrate surface to the molten pool interface with the substrate at its deepest location (nominally along the center line of the bead). The fusion zone depth factors into the remelt ratio (amount of the prior layer(s) remelted by the current pass), and it has been the focus in many previous studies [33][48][50]. The depth model performance is less robust with $R^2 = 0.86$ relative to models of bead height, width, and fusion zone area. The predicted values are more scattered along with the brown diagonal line. Visually, as shown in Fig. 2(c), the molten pool profile is clearly distinguished from the base plate material. A heat affected zone can also be observed in the peripheral area of the molten pool. Compared to the easy measurement of bead height and bead width, the molten pool – heat affected zone interface for fusion zone depth measurement, is a more subjective measurement. This increases measurement uncertainty and can be a major factor for lower prediction performance.

The fusion zone area is the total cross-sectional area normal to the travel direction. In general, it strongly depends on the process parameters such as wire feed speed and robot travel speed. Fusion zone area multiply by the robot travel speed reflects the volume of material deposited per unit time, which is a cost or economic factor consideration during AM process [51]. In Fig. 8(d), the fusion zone area model performs the best among four bead geometry models, as indicated by $R^2=0.97$. The MAE and mean error $\bar{\varepsilon}$ are 5.03 mm$^2$ and 0.08. It also shows the sharpest error histogram with fewer data points sitting outside of the ideal diagonal band.

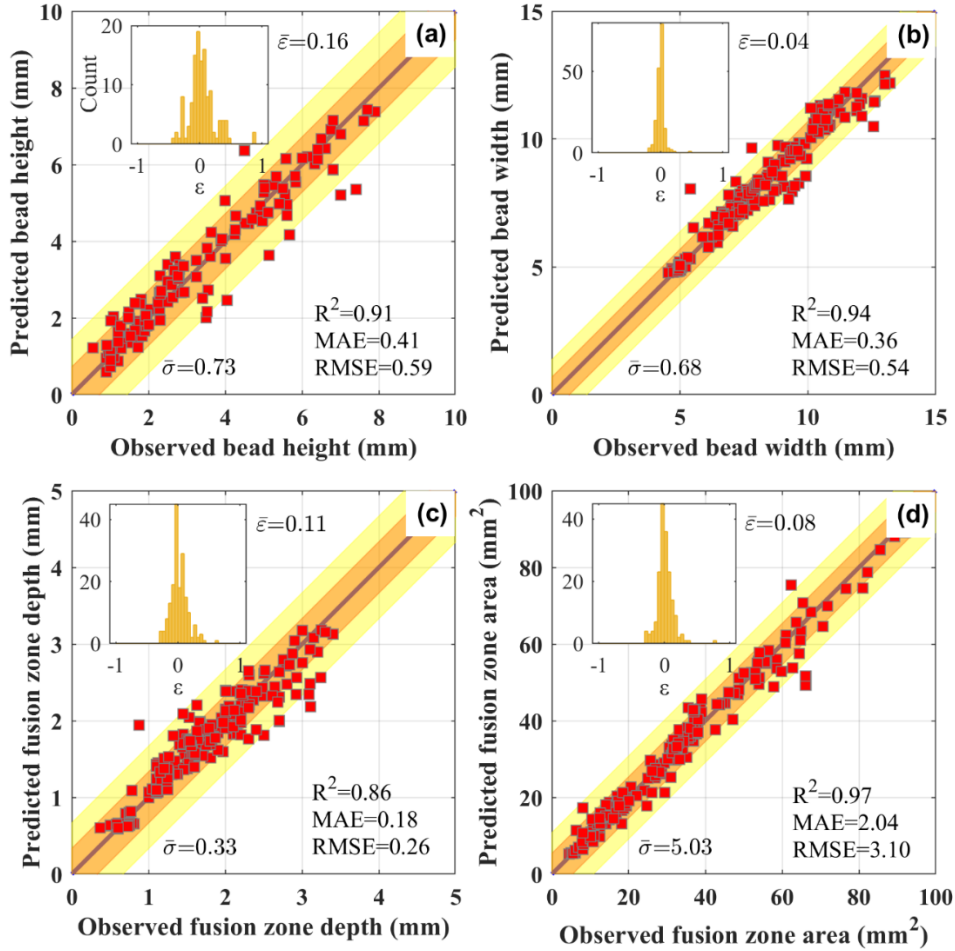

**Figure 8.** Model predictions vs. observations for (a) bead height, (b) bead width, (c) fusion zone depth, and (d) fusion zone area. The inset histogram indicates the relative predicted error ε of all data points. Uncertainty is represented by bands colored at ±σ (orange) and ±2σ (yellow) about the theoretically perfect predicted vs. observed trend line (brown), where σ is the standard deviation of prediction. $R^2$, MAE, RMSE, $\bar{\varepsilon}$, and mean uncertainty $\bar{\sigma}$ values are also shown.

### 3.3.3 Bead microstructure models

The microstructure of Ti-6Al-4V alloy is sensitive to the thermal environment of the process. Different thermal conditions lead to different $\alpha/\beta$ microstructural features and length scales [52][53]. In the case of laser-based AM, as-deposited parts show a transformed $\beta$ microstructure with highly columnar $\beta$ grain (aligned to the build direction) and a fine $\alpha$ lath microstructure [18][54][55]. It has been shown that the final microstructures of Ti-6Al-4V significantly affects mechanical property such as strength, ductility, hardness, fatigue and wear resistance [8][54]. As shown in Fig. 9(a), the $\alpha$ lath thickness is in the range 1.0 to 1.6 µm which is consistent with the previous quantification on the width of $\alpha$ lath thickness [8][54]. The $R^2$ value is 0.83, and mean error $\bar{\varepsilon}$ is only 0.03, as indicated in the text box. The model prediction on $\alpha$ thickness works well and the prediction vs. observed data points follow diagonal line closely.

Fig. 9(b, c) predicts the $\beta$ grain length parallel (para.) to and perpendicular (perp.) to the laser scan direction, respectively. As expected, $\beta$ grains length in the vertical direction (250-2500 µm) are longer than horizontal direction (100-800 µm) due to the preferred growth direction of the $\beta$ phase relative to the dominant thermal gradient. The model prediction on $\beta$ grain length in the parallel plane performs well with $R^2$ =0.95 with a sharper error histogram in the inset $\bar{\varepsilon}$ =0.07, and with mean predictive uncertainty $\bar{\sigma}$ =39 µm. The prediction of $\beta$ grain length in the perpendicular plane with $R^2$ =0.77 is worse than the parallel plane. The predicted $\bar{\varepsilon}$ and uncertainty $\bar{\sigma}$ are also larger (0.15 and 218.40 µm, respectively). The reason could be the length of $\beta$ grain in a vertical direction is difficult to determine due to the bottom of $\beta$ grain is mixed with finer equiaxed $\alpha/\beta$ microstructures as indicated in Fig. 2(c). Another reason is the data problem in the experiment process. There are two noticeable outliers away from the desired diagonal trend when the grain size is large than 2000 µm. Even including the outliers, the prediction of $\beta$ grains length either in a perpendicular or parallel direction is acceptable considering the wide range of length scale and most of the data points prediction agrees with experimental observation.

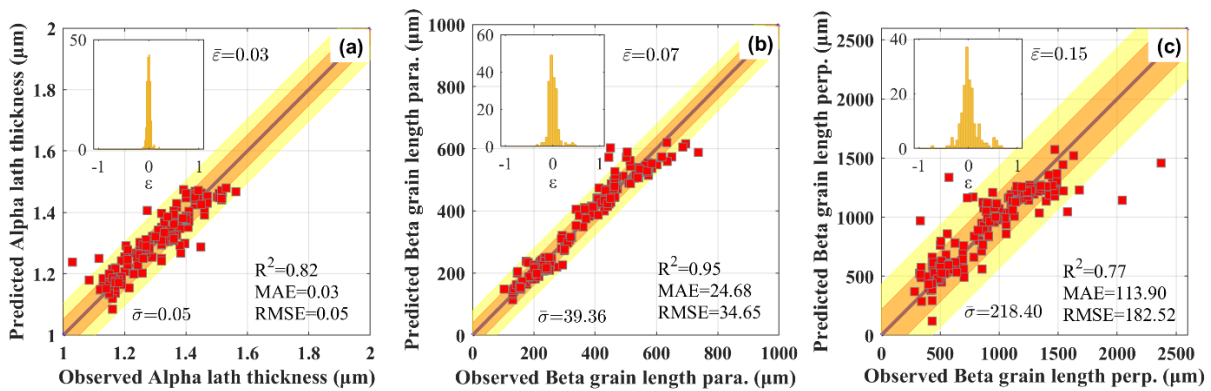

**Figure 9.** Model predictions vs. observations for (a) $\alpha$ lath thickness, (b) $\beta$ grain length parallel to scan direction, and (c) $\beta$ grain length perpendicular to scan direction. The inset histogram indicates relative predicted error ε. Uncertainty is represented by bands colored at ±σ (orange) and ±2σ (yellow) about the theoretically perfect predicted vs. observed trend line (brown), where σ is the standard deviation of prediction. $R^2$, MAE, RMSE, mean relative error $\bar{\varepsilon}$, and mean uncertainty $\bar{\sigma}$ values of predictions are also shown.

To verify the predictive performance of regression models of bead geometry shape and microstructural features, ten-fold cross-validation (CV) was carried out for regression models. The original datasets are split into 10 subsets. The ML model was trained on the training dataset, to test the prediction accuracy of the testing dataset by comparing predictions to the actual observed values. This process is repeated until all subsets have been left out once as testing and their prediction accuracy are calculated. It is interpreted as the ability of the model to generalize the predictions to unobserved process combinations. The testing results can then be averaged to obtain the evaluation metrics, as shown in Table S2. All of the models exhibit a mean relative error $\bar{\varepsilon}$ of less than 0.2. The MAE, RMSE, and uncertainty $\bar{\sigma}$ are low enough considering the property variation range. The $R^2$ value is acceptable except for fusion zone depth and $\beta$ grain length in the perpendicular direction due to uncertainty in the experimental measurements. Generally, the cross-validated models are adequate and acceptable because this magnitude of error is comparable to errors of experimental instrument capabilities and human labeling uncertainties. Once the models are trained properly for capturing these process-property relations across a large process parameter range, the data-driven ML models provide a computationally efficient way for quick property predictions, within a degree of confidence, for process optimization purposes.

### 3.4 Process – property relationship investigation and insight

### 3.4.1 Overall bead quality

Overall bead quality prediction based on the NB classification model is in Fig. 10(a). The bead quality varies with respect to LP, RTS, and WFS. For better visualization and analysis of the 3D map in Fig. 10(a), Fig. 10(b-d) show cross-section planes of the cube at wire feed speed 50, 85, and 100 mm/s. When the space point is predicted one quality label with a confidence value large than 1/3, the bead quality was assigned to this quality label and mapped it to the 3D space in a corresponding color. On the right of Fig. 10(b-d), is shown the predicted confidence of bead quality label that is assigned. It shows that the predicted confidence varies from 0.33 to about 0.9. The confidence value on the label region boundary is lower since each class has equivalent probability at the classification hyperplane region.

The incident laser energy ($E$) is widely used in the literature when evaluating the effects of input energy on the property of the printed part. It allows for the comprehensive evaluation of the range of the process parameters for the desirable property, which is calculated as,

$$E = \frac{LP}{RTS * (\pi r_l^2)} \tag{4}$$

where $r_l$ indicates the laser spot radius. However, it may not necessarily be the best indicator for describing the underlying physics of printing process. For example, WFS and wire pre-heat power are also significant factors for laser energy absorbed. The laser energy absorption coefficient $\eta$ will vary during the build, including the temperature and geometry dependent absorptivity, the shape and phase of the material [56]. Corresponding, the wire materials deposition rate can be calculated by,

$$m = \pi \rho r_w^2 * WFS \tag{5}$$

where $m$ is the mass of wire materials supplied per unit time, $\rho$ is Ti-6Al-4V theoretical density of 4.429 g/cm³, and $r_w$ indicates the wire radius 0.795 mm.

Fig. 10(a) shows that the WFS and RTS are the dominant process settings that predict overall bead quality. When the WFS decreases to below about 80 mm/s, the bead is more likely to have "failed" or "rippled" quality depending on the value of RTS. The "smooth" bead tends to be printed at higher LP. From this, we see that when WFS is set to a higher value (above 80 mm/s), the homogeneous and smooth bead can be built with higher predicted confidence in our prediction space LP 2500 – 6500 W and RTS 1 – 20 mm/s. This is partly because, with higher WFS, more heat was input into the molten pool due to the wire pre-heat power. This is consistent with the previous study on WALM process studies [3][57]. Another reason is the laser energy input; even the lower limit is enough to melt the materials supplied, as indicated by Eq. (4) and Eq. (5). The wire deposition rate $m$ is in the range of 3.5 g/s to 5.25 g/s when the WFS is set between 100 mm/s and 150 mm/s. Considering laser spot focus radius $r_l$ equals wire radius $r_w$, the lower bound of $E$ can be calculated as 63.15 J/mm$^3$ at LP 2500 W and RTS 20 mm/s, which is still in a desirable $E$ range of about 50-100 J/mm$^3$ for Ti-6Al-4V LPBF AM processes [40].

As indicated in the confidence prediction presented in Fig. 10(b,c), the predicted probability of the build quality is not homogeneous. The higher LP has a higher probability of building a "smooth" bead compared to the lower LP. Further, the higher WFS, the more wire material is supplied, the higher laser power is needed to succeed for smooth bead presented by LP – WFS plane. The bead quality at higher WFS seems independent of RTS. As in the right of Fig. 10(b,c), the variation of RTS or LP can also change the probability of bead label correspondingly. At WFS below about 80 mm/s, when it increases RTS, the quality of bead turns from "rippled" to "failed". In this case, the increase of RTS decreases the heat input per unit length deposited on the materials, which is insufficient for a smooth, uninterrupted bead. Besides, combined with this lower WFS, the cooling of molten pool temperature was encouraged to form an intermittent and uneven bead, which leads to failed beads. Conversely, the low RTS provides enough heat input for the molten pool to melt raw materials but insufficient material is fed; thus, forming a rippled bead.

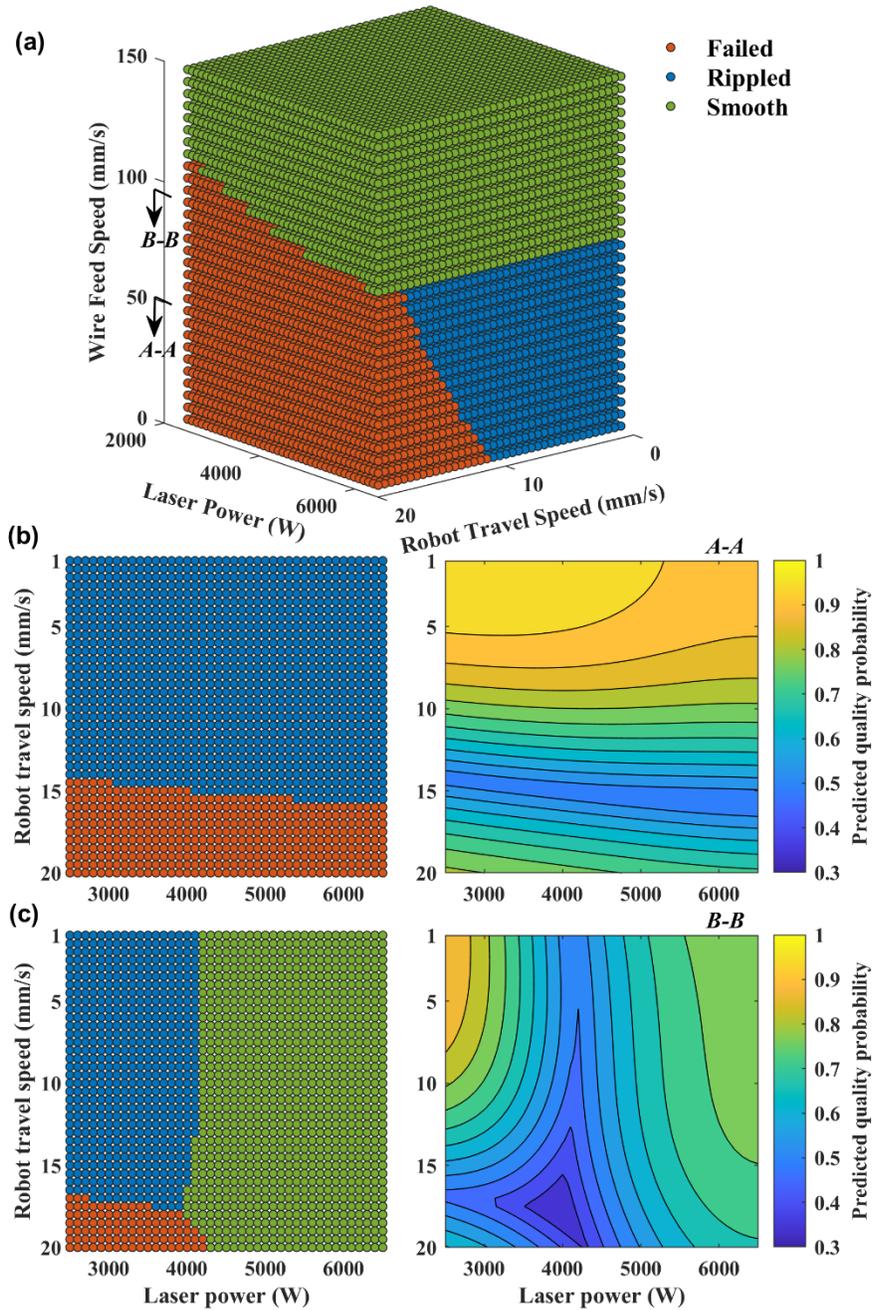

**Figure 10.** (a) Bead overall quality prediction with respect to laser power, robot travel speed and wire feed speed. The cross-section of 3D cube prediction at wire feed speed (b) 50 mm/s and (d) 95 mm/s plane. The contour plot on the right of each plane indicates quality prediction confidence. The 3D contour plot of predicted probability is presented in Figure S1.

### 3.4.2 Bead geometric shape

The prediction of overall bead quality is based on visual appearance without destructive measurements of the sample. Internal features of the bead are also critical as they determine essential aspects of how the beads stack and how multi-bead deposits can be fabricated without major defects or voids. [3][12][58]. The bead geometry parameters, such as bead height, bead width, fusion zone depth, and fusion zone area, are analyzed against the variation of WFS, RTS, and LP, as shown in Fig. 11.

First, the influence of RTS on characteristic bead geometry features is significant, as indicated in Fig. 11(a-d). It is observed that with increasing RTS, the bead height, bead width, fusion zone depth, and fusion zone area decrease accordingly. The general trend is that the specific bead geometry feature is inversely proportional to the RTS. This may be attributed to the more thermal energy is accumulated in the molten pool and resulting in a higher dimension of the printed bead such as bead height, bead width and fusion zone area. However, the increase of geometry feature value is not linear, which is highly dependent on the range of other process parameters. In contrast, earlier publications mostly reported the near-linear fashion of RTS dependence of bead width and depth due to the limited number of data collection [17][49]. For example, in our predicted colored 3D map, at different machine settings of WFS, it is observed that the variation gradients of bead height are not consistent. The bead height is sharply increased at higher WFS, and keeps almost constant at lower WFS below 50mm/s. This is relevant to a large number of input process parameters and is a result of the complex intercorrelation among these parameters.

Laser power is another dominant factor for the characteristic of a single bead. As expected, the bead width, fusion zone depth and fusion zone area are generally proportional to LP, given the fact that more energy is being applied to the molten pool. Specifically, the bead width near-linearly increases from 4 to 11 mm, with the LP increasing to 6500 W. The fusion zone depth increases with the LP with various increment gradients at different WFS and RTS ranges. For example, fusion zone depth increases rapidly with LP in the region of WFS>50 mm/s. Similarly, the LP increases the fusion zone area at the higher WFS region (>100 mm/s). A larger fusion zone area is found with higher WFS, lower RTS, and higher LP, as expected. Interestingly, LP has a different effect on the relations with bead height. The variation of LP is inversely proportional to bead height. The bead height decreases as the LP is increased at a low RTS region. The possible reason is that high heat input per distance and unit time causes a higher molten pool temperature. Liquid surface tension and melt viscosity decrease as temperature increases [59][60]. The wetting angle between the bead surface and substrate is decreased in the range of 0 to 90$^o$. Liquid metal tends to flow and spread due to the reduced resistance and increased driving force [3][17]. Therefore, the bead height is reduced due to the over-heating effects. The fusion zone depth and bead width increase accordingly given the higher energy input, which allows the molten pool to melt deeper/wider into the underlying materials.

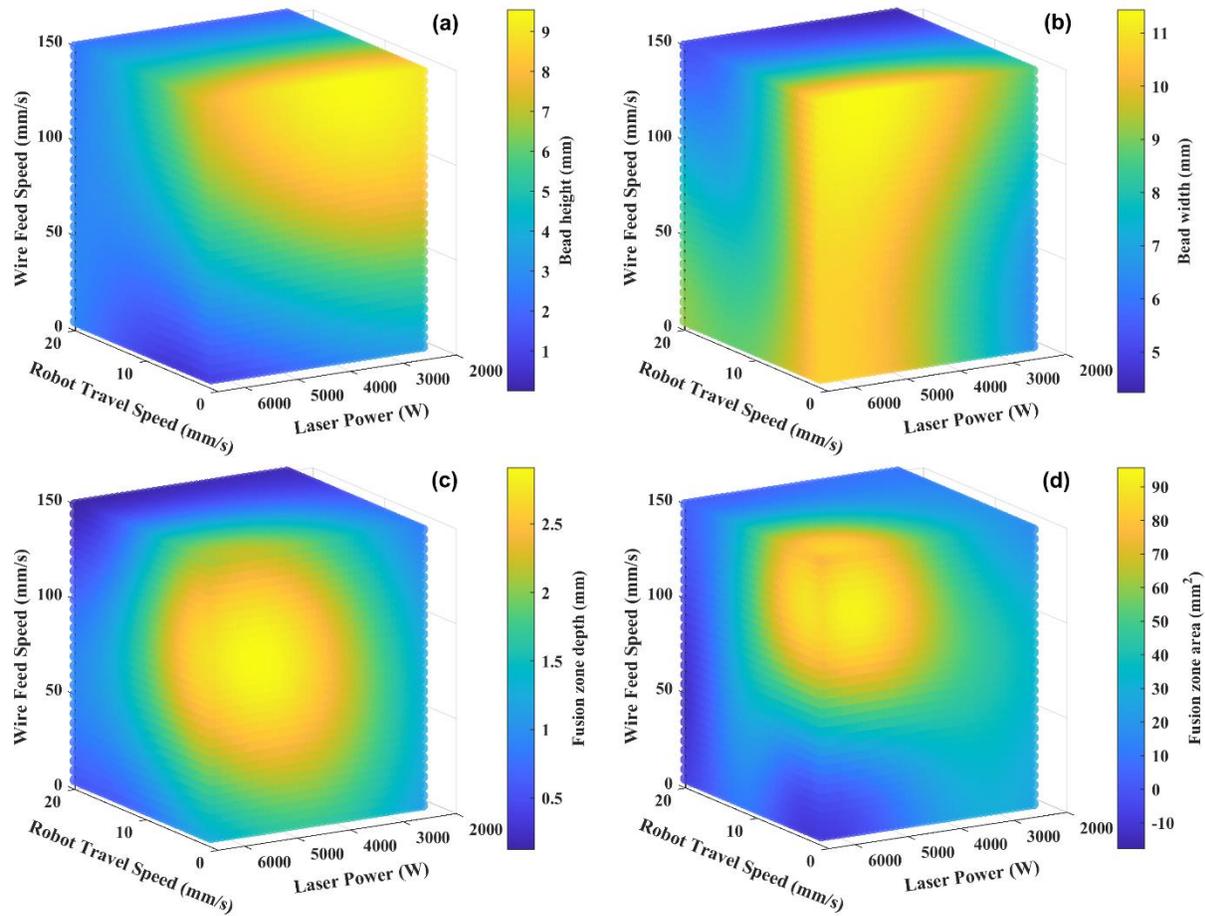

**Figure 11.** The 3D contour representation of bead geometry property prediction on (a) bead height, (b) bead width, (c) fusion zone depth, and (d) fusion zone area. The 3D contour plot of predicted uncertainty $\sigma$ is presented in Figure S2.

For wire feed speed, it exhibits complex relations with the process parameters. With the increase of WFS at a low RTS, bead height and fusion zone area are generally increased. The bead height increases more sharply at lower LP, whereas the fusion zone area increases rapidly at higher LP. Further, it shows that the bead width is not sensitive through the variation range of WFS. That can be explained by that at a given specific RTS and LP, the heat input per unit length and time applied on the molten pool tends to remain constant level; thus, the deposited material has a taller bead as opposed to a wider bead as WFS increases. For fusion zone depth, the WFS is not linearly aligned with depth variations. The fusion zone depth presents the largest value at low RTS and high LP as expected due to the higher laser energy input. But the increment trends are not consistent in that the depth firstly increases then decreases to a small degree with the WFS increasing. This complicated relationship is due to the multiple input process parameters and their complex interaction.

### 3.4.3 Bead microstructures

The microstructure of the deposited material strongly depends on the molten pool temperature, cooling rates, and thermal gradients, which is a function of the machine processing parameters. Understanding microstructure evolution is critical for both process control and process design because it plays an essential role in controlling mechanical properties.

The microstructure of the additively manufactured Ti-6Al-4V part exhibits directional columnar $\beta$ grains parallel with the build direction and very fine lath $\alpha$ [18][54]. Fig. 12(a) shows the prediction on $\alpha$ lath thickness range from 1 μm to 1.6 μm with the variation of process parameters. In general, the $\alpha$ lath thickness is directly related to the cooling rate with faster cooling rates yielding thinner laths. Coarser $\alpha$ phase thickness likely occurs at high WFS and low RTS with relatively low dependence with LP. The $\alpha$ lath thickness tends to decrease at higher RTS due to the faster cooling rate of the molten pool. The finer $\alpha$ phase can be achieved with low molten pool temperature and this higher molten pool cooling rates. On the contrary, coarser $\alpha$ microstructures can be achieved with higher molten pool temperature resulting from a decreased RTS given higher WFS with enough materials supplied [18][61].

While many properties in Ti-6Al-4V are controlled by the microstructure of the $\alpha$ phase, certain properties (particularly fatigue response) can be greatly affected by features of the prior $\beta$ grains, which, in turn, are controlled by phenomena that occur during solidification [62]. Fig. 12(b) shows the prediction of $\beta$ grains length parallel to the travel direction. The prediction of length ranges from about 50 to 850 μm. Except for the highest value at low RTS and high WFS region, most of the $\beta$ grain length range from 50 to 400 μm. Similar effects with $\alpha$ lath thickness can be observed for $\beta$ grain length in which the length decreases with the RTS but remains insensitive with LP, given at the high WFS is providing sufficient feedstock into the molten pool. This is consistent with an earlier publication in laser AM Ti-6Al-4V that addressed how the variation of the laser power and transverse speed can influence the width of the prior $\beta$ grains [62]. They observed that the $\beta$ grain width decreases with traverse speed but remains unaffected by the laser power. This is expected given that the grain size tends to decrease with the cooling rate, which arises from a high RTS.

Fig. 12(c) represents the prediction of $\beta$ grain length in the perpendicular direction to the laser scan direction (parallel to the build direction). The predictive $\beta$ grains length in this direction is significantly larger than the horizontal direction due to the direction of the thermal gradient and the preferred growth orientation of the $\beta$ phase. The relationship of process parameters to $\beta$ grain length in the perpendicular direction is more complex relative to other properties. The higher $\beta$ grain length at the low RTS region makes sense due to the higher incident laser energy. However, it also exhibits the hot spot region at high RTS, high WFS, and low LP machine settings. This is all primarily driven by the thermal gradient (it is steepest in the build direction) and the fact that the solidifying $\beta$ phase is body centered cubic crystal structure and this drives a very columnar grain growth (with ⟨001⟩ direction favored). At high WFS, the injected materials directly adding up together to form a very high bead. The accumulated heat from LP and pre-heat wire power is conducted to the cold substrate, a faster cooling rate is obtained in the build direction. Therefore, $\beta$ grains length tends to enlarge and grow in the directional direction. Even this explanation and insight may work, we should notice that the predicted uncertainty is also relatively larger for $\beta$ grains length in a perpendicular direction, as indicated in Fig. S3(c). The large predictive uncertainty may source from experimental measurements error, as discussed before. The considerable uncertainty also likely occurred in regions with fewer nearby experimental points, such as boundaries, due to the intrinsic fact of having less neighborhood information to make a prediction and expected extrapolation error [43].

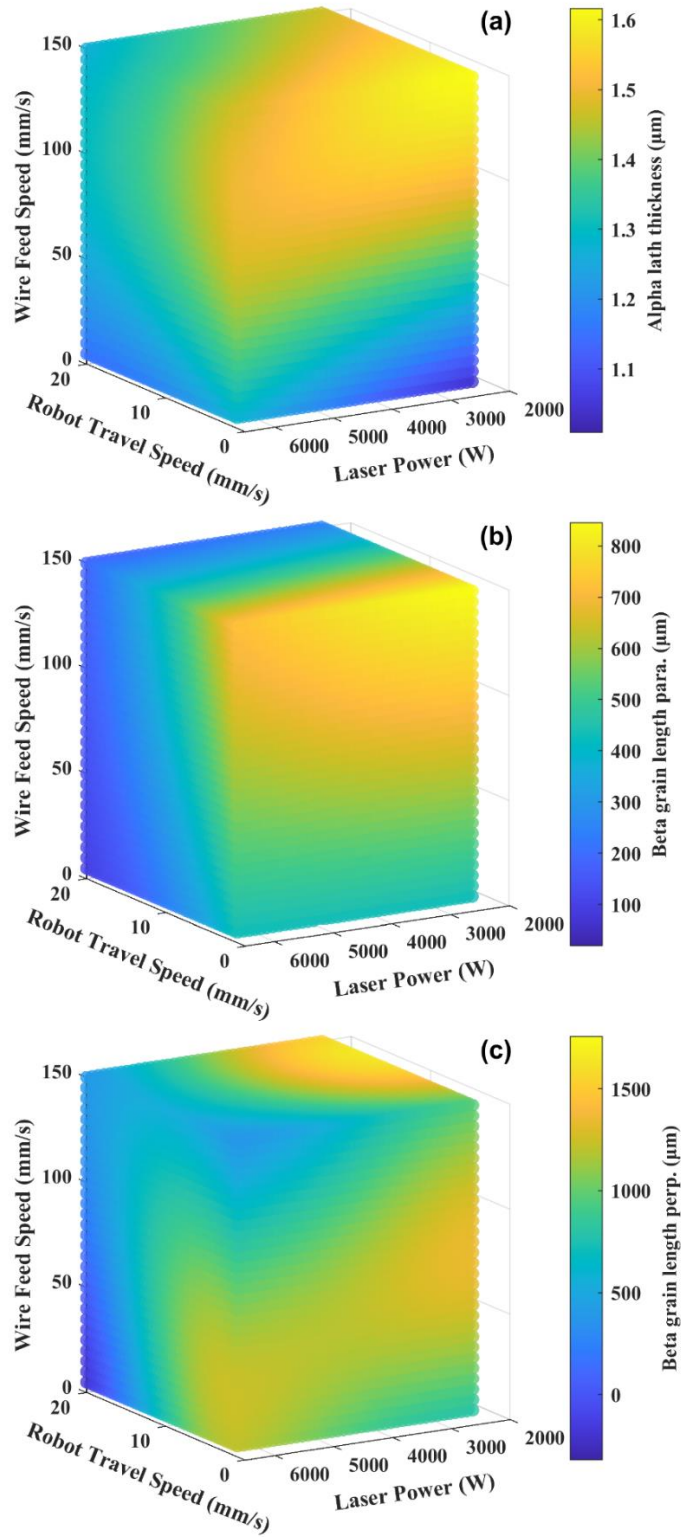

**Figure 12.** The 3D contour representation of prediction on (a) $\alpha$ lath thickness, (b) $\beta$ grain length in parallel to scan direction, and (c) $\beta$ grain length in perpendicular to scan direction. The 3D contour plot of predicted uncertainty $\sigma$ is presented in Figure S3.

### 3.4.4 Bead dilution and aspect ratio

As discussed in Fig. 11 for individual bead geometry prediction, there is the geometry shape coupling effect with the process parameters. For example, as the increase of laser power, the bead width and fusion zone depth tend to increase while the bead height is depressed due to the lowered surface tension of the molten material as the temperature of the molten pool increase. Therefore, it is critical to explore and visualize a more comprehensive indicator for bead geometry shape. Bead dilution (*Du*) and aspect ratio (*AR*) are some choices, which are expressed as,

$$Du = \frac{d}{h+d} \tag{6}$$

$$AR = \frac{w}{h} \tag{7}$$

where *d* is the fusion zone depth or penetration depth of printed bead into the substrate. *h* is the height of the material deposited above the substrate and *w* is the width of bead at the bead-substrate interface. The more comprehensive geometry indicator bead *Du* and *AR* are subsequently defined and analyzed in Fig. 13.

A lower *Du* corresponds with a lower laser power or a larger amount of feedstock material which acts to quench the molten pool. On the contrary, higher power correlates with lower bead printing height relative to fusion zone penetration. Similarly, *AR* represents the ratio of bead width and bead height. Low AR leads to taller beads while higher AR leads to flatter beads.

The prediction of bead geometry features including bead height, width, and fusion zone depth in Fig. 11, derives the more comprehensive geometry property *Du* and *AR*. Consequently, it can be used to get insight and identify combinations of process parameters within either too high or too low *Du* and *AR* that might negatively affect bead quality. Fig. 13(a, b) displays *Du* and *AR* prediction across process parameters in 3D contour map. The supplementary data are the original figures in 3D view, that can be readily used to identify the *Du* and *AR* variations relative to processing parameters. It observes the *Du* relations with process parameters are in more complex nonlinear relations than individual bead geometry prediction. *A-A*, *B-B*, and *C-C* in the LP-RTS plane indicate the cross-section of 3D prediction cube to better visualize the cube inside. It shows the prediction of *Du* is in the range from about 0.06 to 0.58. The hot regions with the highest *Du* (>0.4) are highlighted at the center region of LP-RTS plane and low WFS, being >0.4. At other process regions WFS>50 mm/s, the *Du* is ranged from 0.06 to 0.4. The hot region of *Du* is more likely clustered at low WFS. This is reasonable since it represents the desirable laser energy is applied and limited wire material is supplied. Thus, the fusion zone depth is larger than the bead height. Another interesting phenomenon captured is that with WFS increasing from *A-A* to *C-C* plane, the highest *Du* hot region at the LP-RTS plane is shifted to a higher laser power range. This indicates that higher LP is needed for enough fusion zone penetration when the wire materials supplied are increased.

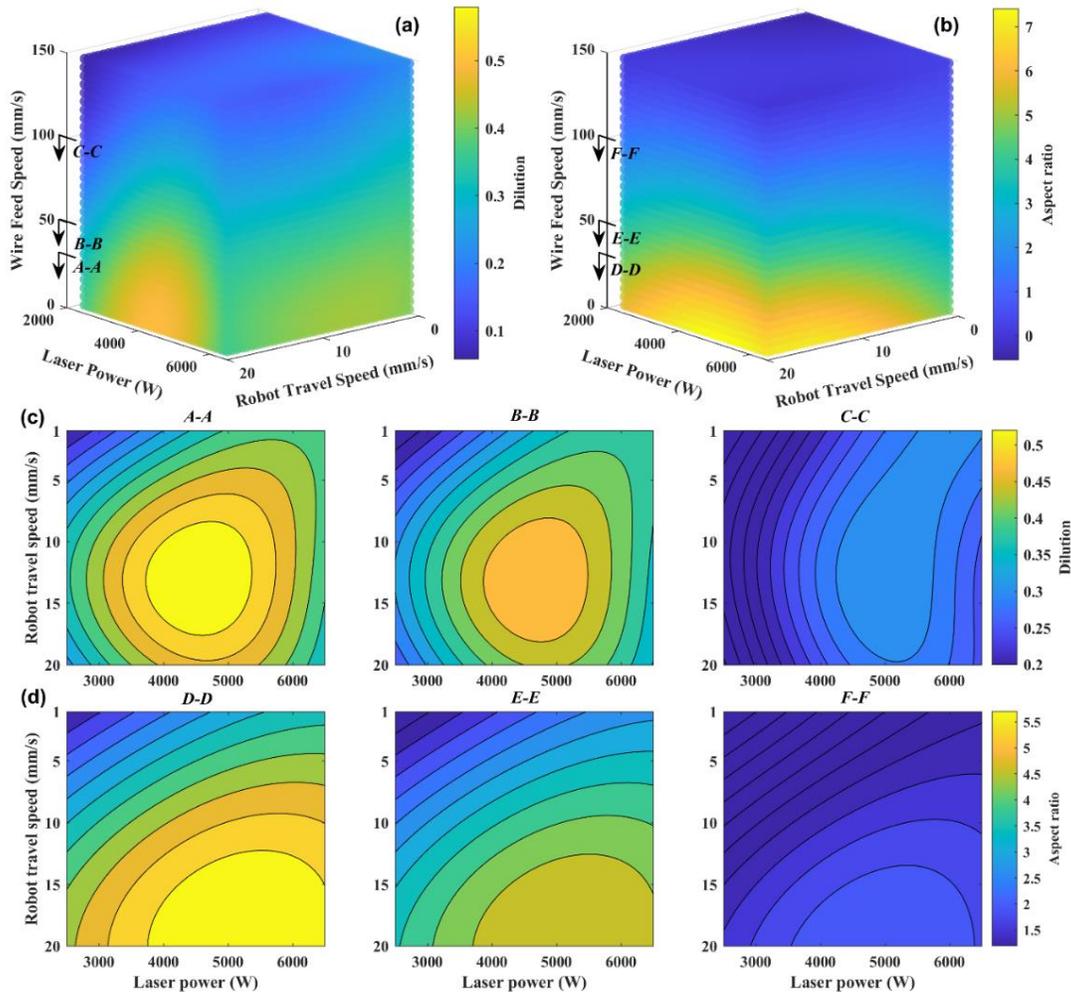

**Figure 13.** The 3D contour representation of prediction on (a) dilution, and (b) aspect ratio. (c) The 2D contour plot representation of cross-section of (a) at wire feed speed 30 mm/s, 50 mm/s

and 100 mm/s. (d) The 2D contour plot representation of cross-section of (b) at wire feed speed 30 mm/s, 50 mm/s and 100 mm/s. The 3D contour plot of predicted uncertainty $\sigma$ is presented in Figure S4.

Fig. 13(b) represents the prediction of the bead aspect ratio relative to process parameters. A similar trend as bead dilution is exhibited at 3D cube and cross-section plane *D-D*, *E-E*, and *F-F*. It shows the predictions of Du are up to 7.4. The highest *AR* region (>4) is at the bottom corner of the 3D cube (low WFS, high LP and high RTS), due to too enough energy input applied, and the material is fused completely to spread and form a flat bead shape. At other process regions, the *AR* is more likely lower between 0 to 4. The hot region of *Du* is more likely clustered at low WFS.

### 3.4.5 Wire heat power effects

The wire heat power (WHP) is another source of input power into the system in addition to the laser power. The WHP pre-heats the wire in the feed tube such that it enters the molten pool near the melting point of the alloy. This heat imparted to the wire feedstock can reduce the amount of heat needed from the laser, which can lead to a refinement in the typical columnar grains in Ti-based alloys [63]. As the results in Sections 3.1 & 3.2, the WHP is also an important and adjustable process parameter for determining the characteristics and quality of the bead. The WHP ranges from 0 to 1.2 kW in the experimental data collection, as shown in Fig. 3. Previous predictions in Figs. (8-11) are performed at a constant WHP value 0.6 kW. In this part, the predictions are performed against the variations of WHP in the range of 0-1.5 kW, RTS (1-20 mm/s) and LP (2500-6500 W), in which a constant WFS=95 mm/s is used to assess the influence of WHP on various bead properties.

Fig. 14 shows the prediction of bead overall quality as a function of WHP, RTS and LP. The cross-section of bead quality 3D cube is generally similar to the RTS-LP prediction plane of WFS=85 mm/s and WHP=0.6kW, in Fig. 10(c). The failed bead is obtained at high RTS and low LP regions due to the insufficient laser energy input. The smooth bead is more likely deposited when operating at high LP. The WHP can have a direct impact on the overall bead quality. For example, with the increase of WHP, the boundary between smooth and rippled in the WHP-LP plane shifts to a lower LP due to the additional energy input from wire pre-heat power.

Fig. 15 presents the bead geometry property variations at WFS=95 mm/s. In the vertical WHP plane, with the increase of WHP, the bead geometry features such as bead height, fusion zone depth and area are generally increased. However, from the heat contour in the WHP-LP plane of Fig. 15(b), it seems the bead width is complementary to that of bead height prediction in Fig. 15(a). The underlying reason may come from the coupling effects of bead geometry characteristics. For example, the lower bead height is prone to be with wider/higher bead width (at low RTS and high LP), because the molten pool is more likely to spread at high laser energy input and vice versa.

In addition, the prediction of microstructures against the variations of WHP, RTS and LP are presented in Fig. S5, in which the $\alpha$ lath thickness and $\beta$ grain length in parallel direction exhibit little dependence of WHP by observing from WHP vertical plane. In the WHP-LP plane of Fig.

S5(c), it shows $\beta$ grain length in perpendicular direction decreases as WHP is increased. This is mainly because of the refining effects of wire pre-heating, resulting in a mixture of short columnar grains and equiaxed ones [63][64]. The predictions of bead dilution and aspect ratio are shown in Fig. S6. From the vertical WHP plane in Fig. S6(a), it seems the bead dilution increases and then decreases as the WHP is increased from 0 to 1.5 kW. From WHP-RTS plane of Fig. S6(b), the aspect ratio keeps almost constant and then decreases as the WHP is increased up to 1.0 kW.

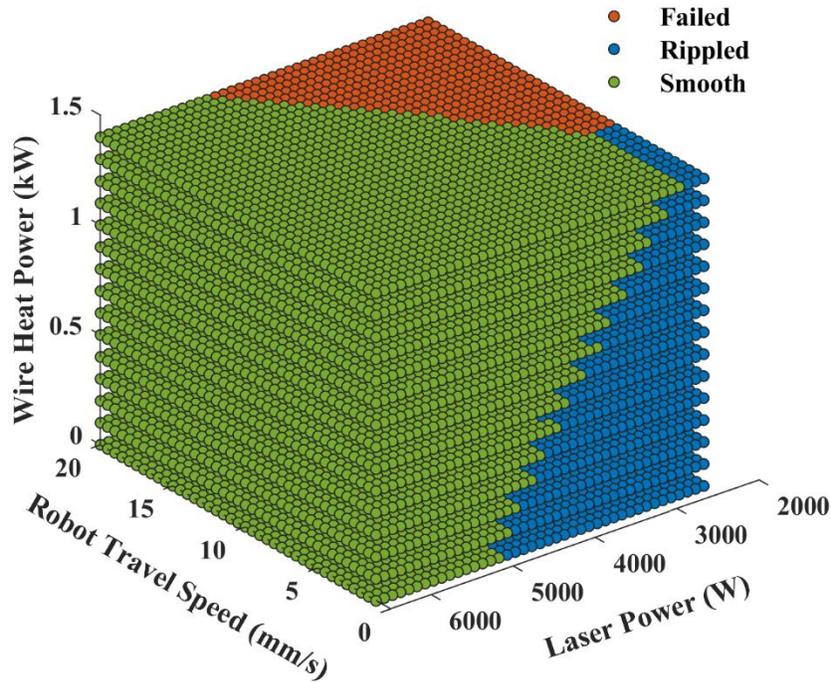

**Figure 14.** The effects of wire heat power. Bead overall quality prediction with respect to wire heat power, laser power and robot travel speed at constant wire feed speed 95 mm/s.

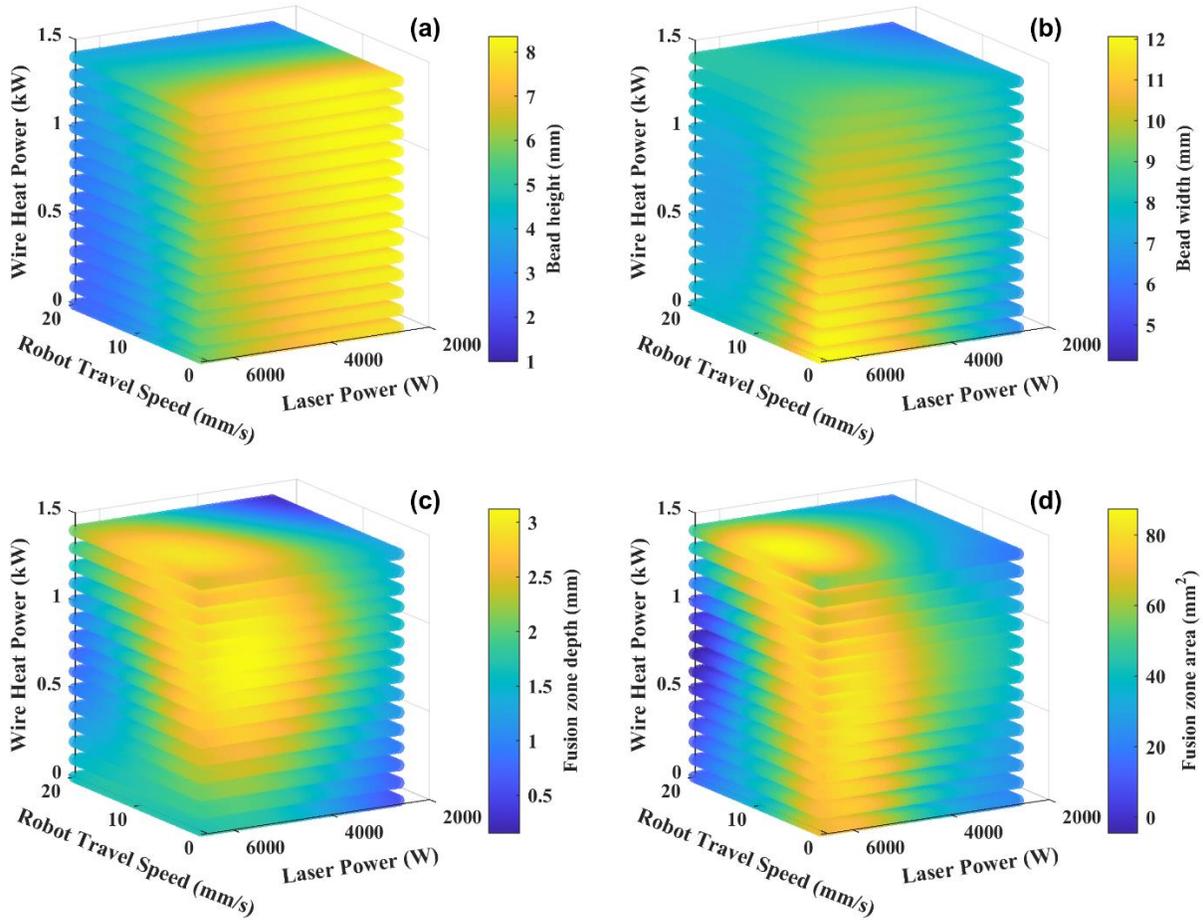

**Figure 15.** The effects of wire heat power. Bead geometric shape prediction with respect to wire heat power, laser power and robot travel speed at constant wire feed speed 95 mm/s.

## 4. Conclusions

In this article, a comprehensive quality material design framework is presented based on machine learning from the experimental data of WLAM Ti-6Al-4V. It is shown to predict their performances in a high dimensional, multiple-target-property design space that considers multi-step AM processing routes, and characterization property variations. The results highlight that the influences and correlations of process variables on the property of printed beads are comprehensively investigated. It provides an effective method for assisting in the selection of the optimum process parameters for the desired outcome.

1) A relatively comprehensive database is collected that consists of multiple processing variables under a set of controlled experiments. The characterization property includes the overall bead quality label, bead cross-section geometry features, and microstructural features.

2) The correlations of processing parameters with the property are analyzed with mutual information score and Pearson matrix. The process – microstructure – geometry relations are modeled with data-driven ML models for revealing the hidden complex relationships. The models are rigorously optimized, tested with error metrics and cross-validation.

3) The property prediction methodology is proposed to predict the bead overall quality, geometric shape, and microstructural properties across the designed processing space in a 3D contour plot. The insights and effects of process variables on the characteristics of bead are discussed thoroughly as a process – property relations interpretation.

Finally, even the case study is demonstrated on the additively manufactured Ti-6Al-4V alloy, the developed material design modeling and prediction methodology are more general. It can be extensible to other material systems with multiple properties that are of interest, such as high-entropy alloys, steels, ceramics, and semiconductor material, given an accumulated database available, no matter the database is in experiments, mechanics-simulations, prior published data, or mixed form.

**Availability of data and material**

The raw data required to reproduce the findings of this work currently can only be shared with the permission of the U.S. Office of Naval Research.

**Acknowledgments**

None.

**Supplementary materials**

See attached file for supplementary materials.